\documentclass[letterpaper,12pt]{article}
\usepackage{fullpage}
\usepackage{amsmath, amsthm, amssymb}
\usepackage{graphicx}
\usepackage{verbatim}
\usepackage{fancyhdr}
\usepackage[colorlinks=true,linkcolor=blue]{hyperref}
\usepackage[all]{xy}

\theoremstyle{plain}
\theoremstyle{definition}

\graphicspath{{Images/}}


\title{\textbf{Analysis of hashrate-based double-spending}}

\author{Meni Rosenfeld\footnote{The author can be contacted at meni@bitcoil.co.il. If you would like to support this research, you can send bitcoins to 17aZjvq7oDgK3Qw1SVtt89mMjEQx4mDoVk.}\\
}

\date{December 11, 2012\\
Latest version: \today}

\begin{document}

\maketitle
\begin{abstract}
Bitcoin (\cite{Bitcoin}) is the world's first decentralized digital currency. Its main technical innovation is the use of a blockchain and hash-based proof of work to synchronize transactions and prevent double-spending the currency. While the qualitative nature of this system is well understood, there is widespread confusion about its quantitative aspects and how they relate to attack vectors and their countermeasures. In this paper we take a look at the stochastic processes underlying typical attacks and their resulting probabilities of success.

\end{abstract}
\pagenumbering{roman}
\tableofcontents
\section{Introduction}\label{chap:intro}
\pagenumbering{arabic}
The Bitcoin system revolves around the concept of transactions, digitally signed announcements that the owner of some coins agrees to transfer them to a different owner. The sender of coins will typically expect some product or service in return.

This concept will be undermined if the sender would be able, after receiving the product, to broadcast a conflicting transaction sending the same coin back to himself. As long as the recipient cannot be sure the coins are his to stay and that they cannot be redirected to another party (without his consent), he would not be safe to deliver a product in exchange. A double-spending attack is in fact a successful attempt to first convince a merchant that a transaction has been confirmed, and then convince the entire network to accept some other transaction; the merchant would be left with neither product nor coins, and the attacker will get to keep both.

This is a problem of synchronization -- there needs to be some universally accepted signal indicating that some transaction is final and that no conflicting transaction can ever be accepted. Given two conflicting transactions, it does not really matter which of them will be accepted, as long as there is a way to know that one transaction has been accepted and can no longer be reversed.

Bitcoin solves this with a proof-of-work system: Computational effort (consisting in the calculation of hashes) is spent on acknowledging groups of transactions, called blocks; and a transaction is considered final once sufficient work has gone into acknowledging the block that contains it. By linking the blocks to form a chain, the total work spent on any transaction is perpetually increasing, making it difficult to elevate any conflicting transaction to the same confirmation status without a prohibitive computational effort.

However, if the attacker is in fact in control of substantial computational power, he may succeed in doing just that. Satoshi Nakamoto's original Bitcoin whitepaper (\cite{Satoshi}) contains a discussion of the statistical aspects of this problem; in this paper we clarify and expand on this work.

\section{The blockchain and branch selection}
Bitcoin transactions are grouped into blocks. Every block references an earlier block by including the uniquely identifying hash of this earlier block in its header. The one exception is the first ever block, known as the \emph{genesis block}, which of course cannot reference an earlier block.

The blocks hence form a tree, with the genesis block as the root, and each block being a child of the block it references. A branch in this tree is a path from a leaf block to the genesis block; each such branch represents one version of the history of Bitcoin transactions. Each branch must be internally consistent and can never include two conflicting transactions; however, the branches need not be consistent with one another, and one branch can include a transaction which contradicts a transaction in another branch.

Since a single coherent history of transactions is desired, at every point one branch in the tree is considered the valid block chain. It is agreed that every node will consider the longest branch it is aware of as the valid chain (more precisely, the branch representing the most proof of work). If multiple branches are tied, the one of which the node learnt first is considered valid by it, until the tie is broken. When nodes are creating new blocks, they are expected to reference the leaf of the valid branch, extending the chain.

Different nodes may be in temporary disagreement about the valid blockchain, if they learnt of tied branches at different times; this is known as a blockchain fork. However, this is quickly resolved when a new block is found, as including it will make one branch longer than the other, a fact with which all nodes can agree. This is illustrated in \autoref{fig:ETree}.

\begin{figure}[h]
\[
\begin{array}{cccc}
\xymatrix{&{\phantom{\square}}&\\\square\ar@2{->}[dr]&\square\ar@2{->}[d]&\\&\square\ar@2{->}[d]&\\\square\ar@2{->}[dr]&\square\ar@2{->}[d]&\\&\square\ar@2{->}[d]&\\&\square}
&
\xymatrix{&{\phantom{\square}}&\\\square\ar@2{->}[dr]&\square\ar@2{->}[d]&\\&\square\ar@2{->}[d]&\\\blacksquare\ar@2{->}[dr]&\square\ar@2{->}[d]&\\&\blacksquare\ar@2{->}[d]&\\&\blacksquare}
&
\xymatrix{&{\phantom{\square}}&\\\blacksquare\ar@2{->}[dr]&\square\ar@2{->}[d]&\\&\blacksquare\ar@2{->}[d]&\\\square\ar@2{->}[dr]&\blacksquare\ar@2{->}[d]&\\&\blacksquare\ar@2{->}[d]&\\&\blacksquare}
&
\xymatrix{&\blacksquare\ar@2{->}[d]&\\\square\ar@2{->}[dr]&\blacksquare\ar@2{->}[d]&\\&\blacksquare\ar@2{->}[d]&\\\square\ar@2{->}[dr]&\blacksquare\ar@2{->}[d]&\\&\blacksquare\ar@2{->}[d]&\\&\blacksquare}
\\
(a)&(b)&(c)&(d)
\end{array}
\]
	\caption{The block tree and branch selection. (a) A possible structure of the block tree at one point in time. (b) The marked branch is invalid because its length is only 3, while there are longer branches. (c) A branch which is tied for the highest length, and can be considered valid by some nodes. (d) If a new block is found referencing the leaf of the other branch, that branch becomes the longest and is agreed by all nodes as valid.}
	\label{fig:ETree}
\end{figure}

A transaction is said to have $n$ confirmations if it is included in a block which is part of the valid chain, and there are $n$ blocks in the path from that block to the leaf of the chain, inclusive. It is generally assumed that a transaction with enough confirmations is safe from double-spending; in the next sections we will explore the conditions under which this assumption holds.

\section{Playing catch-up}\label{sec:catchup}
A successful double-spending attack consists of the following steps:
\begin{enumerate}
\item{Broadcast to the network a transaction in which the attacked merchant is paid.}
\item{Secretly mine a branch which builds on the latest block at the time (before the transaction made it into a block), which includes instead a conflicting transaction which pays the attacker.}
\item{Wait until the transaction to the merchant receives enough confirmations and the merchant, confident in his payment, sends the product.}
\item{If necessary, continue extending the secret branch (which contradicts the transaction) until it is longer than the public branch (which includes the transaction), then broadcast it. Because the new branch is longer than the one currently known by the network, it will be considered valid, and the payment to the merchant will be replaced by the payment to the attacker.}
\end{enumerate}

The process is illustrated in \autoref{fig:doublespend}.

\begin{figure}[h]
\[
\begin{array}{ccc}
\xymatrix{&{\phantom{\square}}&\\&{\phantom{\square}}&\\&{\phantom{\square}}&\\&\blacksquare\ar@2{->}[d]&\\&\vdots\ar@2{->}[d]&\\&\blacksquare&}
&
\xymatrix{&{\phantom{\square}}&\\\blacksquare\ar@2{->}[d]&&\\\blacksquare\ar@2{->}[dr]&\square\ar@2{->}[d]&\\&\blacksquare\ar@2{->}[d]&\\&\vdots\ar@2{->}[d]&\\&\blacksquare&}
&
\xymatrix{&\blacksquare\ar@2{->}[d]&\\\square\ar@2{->}[d]&\blacksquare\ar@2{->}[d]&\\\square\ar@2{->}[dr]&\blacksquare\ar@2{->}[d]&\\&\blacksquare\ar@2{->}[d]&\\&\vdots\ar@2{->}[d]&\\&\blacksquare&}
\\
(a)&(b)&(c)
\end{array}
\]
	\caption{Outline of double-spending. (a) The state of the block chain when the attack starts. The leaf block does not have any of the relevant transactions yet. (b) The branch on the left is known to the network, and includes the transaction paying the merchant with 2 confirmations. The merchant now sends the product. Meanwhile, the attacker has found 1 block in an alternative private branch which credits himself instead. (c) If the attacker manages to get his branch to be longer than the one known by the network, he releases it and the payment to himself is now accepted by the network.}
	\label{fig:doublespend}
\end{figure}

When attempting a double-spend, the attacker will typically find himself in the following situation: The network currently knows a branch which credits the merchant, which has $n$ blocks on top of the one in which the fork started. The attacker has a branch with only $m$ additional blocks, and both are trying to extend their respective branches. Will he ever be able to catch up with the network and come up with a longer branch? Or will the gap just keep increasing, leaving the attacker behind with a branch which will never be valid?

To model this, we make a few simplifying assumptions which we will use throughout the paper:
\begin{enumerate}
\item{The total hashrate of the honest network and the attacker is constant. Combined they have a hashrate of $H$, of which $pH$ belongs to the honest network and $qH$ belongs to the attacker, where $p+q=1$.}
\item{The mining difficulty is constant, and such that with a hashrate of $H$ the average time to find a block is $T_0$.}
\end{enumerate}

Let us denote by $z=n-m$ the number of blocks by which the honest network has an advantage over the miner. Whenever a block is found, the value of $z$ changes; if that block was found by the honest network $z$ increases by 1, and if that block was found by the attacker $z$ decreases by 1. Formally, this is a continuous-time Markov chain with a rate of $p/T_0$ for moving up a step and a rate of $q/T_0$ for moving down a step. \footnote{By assumption, within $\epsilon$ units of time, a miner with hashrate of $H$ has a probability of $\epsilon/T_0$ to find a block. The honest network, with a hashrate of only $pH$, has a probability of $\epsilon p/T_0$, and so on.}

If $z$ ever reaches $-1$, the attacker's chain becomes longer and his attack succeeds. If this never happens, the attack fails. Because we are interested in \emph{whether} $z$ will ever become $-1$, and not \emph{when} this will happen, this process is equivalent to a discrete-time Markov chain, where each step is defined as finding a block by anyone. It is easy to see that when a block is found, it has a probability of $p$ to be by the honest network and a probability of $q$ to be found by the attacker, and hence this process can be concisely summarized as:

\[z_{i+1}=\left\{\begin{array}{ll}z_i+1&\textrm{with probability $p$}\\z_i-1&\textrm{with probability $q$}\end{array}\right.\]

\begin{figure}[h]
\[
\begin{array}{cc}
\includegraphics[width=80mm]{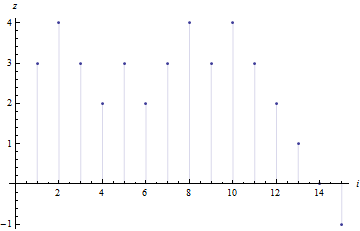}
&
\includegraphics[width=80mm]{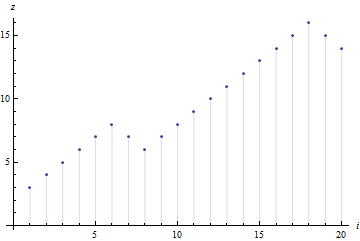}
\\
(a)&(b)
\end{array}
\]
	\caption{Possible scenarios of catching up. (a) A successful double-spend attempt; after 15 blocks were found, the attacker managed to extend his fork to be longer than the network's chain. (b) A failed attempt; within 20 blocks, the network has obtained such a significant advantage that the attacker's chance to ever catch up is negligible.}
	\label{fig:catchup}
\end{figure}

Two possibilities for how this process can unfold, a successful and a failed double-spend attack, are depicted in \autoref{fig:catchup}.

We will denote by $a_z$ the probability that the attacker will be able to catch up when he is currently $z$ blocks behind. Clearly, if $z<0$ then $a_z=1$ as the attacker already has a longer branch. To find $a_z$ when $z\ge0$, we condition on the first step. If the next block found will be by the honest network, which happens with probability $p$, the attacker will now be $z+1$ blocks behind and his probability of success will be $a_{z+1}$. If the next block found will be by the attacker, which happens with probability $q$, his probability of success will be $a_{z-1}$. It follows that $a_z$ satisfies the recurrence relation
\[a_z = pa_{z+1}+qa_{z-1}.\]
It is easy to solve this (noting that $p+q=1$ and the boundary conditions) to find that
\[a_{z}=\min(q/p,1)^{\max(z+1,0)}=\left\{\begin{array}{ll}1&\textrm{if $z<0$ or $q>p$}\\(q/p)^{z+1}&\textrm{if $z\ge0$ and $q\le p$}\end{array}\right.\]
This is only a preliminary result for the analysis of a double-spending attack in its entirety, but a few notable things are already apparent:
\begin{itemize}
\item{The probability of success depends on the number of blocks, and \emph{not} on the time constant $T_0$.}
\item{If the attacker controls more than half of the total network hashrate, he always succeeds in catching up, from any disadvantage.}
\item{When $q<p$, the probability of success decreases exponentially with the disadvantage $z$; the lower $q$ is, the faster the decay.}
\end{itemize}

\section{Waiting for confirmations}
The standard practice for a merchant is to wait for $n$ confirmations of the paying transaction, and then provide the product. While the network is finding these confirming blocks, the attacker is building his own branch which contradicts it. What is the probability that he will successfully double-spend?

Before $n$ confirmations are obtained, the attacker cannot release his fork, even if it is longer, since it will dissuade the merchant from completing the order. He must wait for $n$ confirmations, and then either release his branch if he has the advantage, or continue to work on it hoping that he gains it.

The chances of success crucially depend on his disadvantage, that is, the value of $z$, at the time $n$ confirmations are reached. Satoshi's paper makes the simplifying assumption that $n$ blocks are found by the honest network in the average time, $\frac{nT_0}{p}$, and that accordingly $m$, the number of blocks found by the attacker in this time, follows the Poisson distribution with mean $n\frac{q}{p}$ (despite the mention of time, $T_0$ still has no role in the result). We will not use this assumption, but rather model $m$ more accurately as a negative binomial variable; it is the number of successes (blocks found by the attacker) before $n$ failures (blocks found by the honest network), with a probability $q$ of success. The probability for a given value of $m$ is
\[P(m) = \binom{m+n-1}{m}p^nq^m.\]
Once $n$ blocks are found by the honest network, in a period of time during which $m+1$ blocks are found by the attacker (we assume one block was pre-mined by the attacker before commencing the attack), the race starts as we have analyzed in \autoref{sec:catchup}, with $z=n-m-1$. It follows that the probability for the double-spend to succeed, when the merchant waits for $n$ confirmations, is equal to
\begin{align}
r&=\sum_{m=0}^{\infty}P(m)a_{n-m-1}\nonumber\\
&=\sum_{m=0}^{n-1}\binom{m+n-1}{m}p^nq^m\left(\min(q/p,1)\right)^{n-m} + \sum_{m=n}^{\infty}\binom{m+n-1}{m}p^nq^m\nonumber\\
&=\begin{cases}1-\sum_{m=0}^{n}\binom{m+n-1}{m}(p^nq^m-p^mq^n) & \text{if } q<p \\1&\text{if } q\ge p  \end{cases}\label{eqn:prob}
\end{align}

\section{Graphs and analysis}
\autoref{eqn:prob} is a bit unwieldy, so we will now present a few ways to visualize it.

\begin{figure}[h]
\includegraphics[width=160mm]{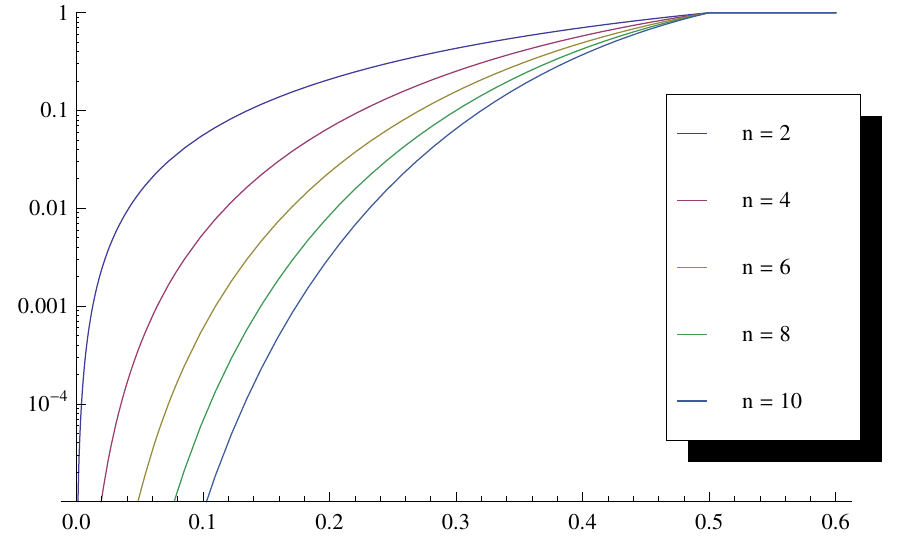}
\caption{The probability $r$ of successful double spend, as a function of the attacker's hashrate $q$, for different values of the number of confirmations $n$. For $q>0.5$, the probability is always 1. The graph is in logarithmic scale; the lowest value shown is $10^{-5}$, or $0.001\%$.}
\label{fig:prob}
\end{figure}

In \autoref{fig:prob}, we see the probability $r$ of successful double spend, as a function of the attacker's hashrate $q$, for different values of the number of confirmations $n$. More confirmations decrease the success rate, but whatever is the number of confirmations, the success rate approaches 100\% as the attackers hashrate approached 50\% of the total network hashrate; and for $q>0.5$, the attack will always succeed.

\begin{figure}[h]
\[
\begin{array}{cc}
\includegraphics[width=80mm]{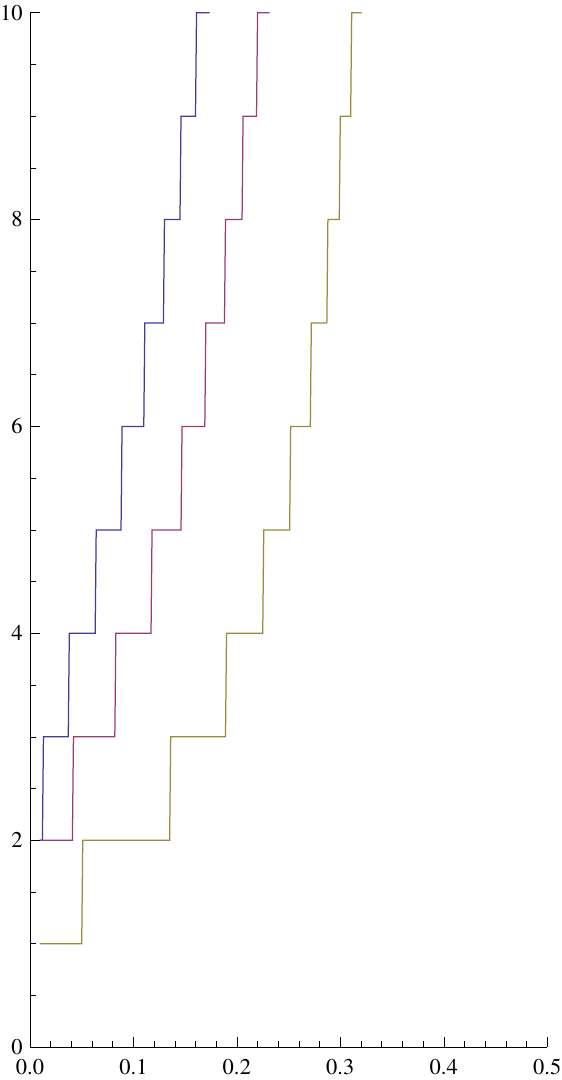}&
\includegraphics[width=80mm]{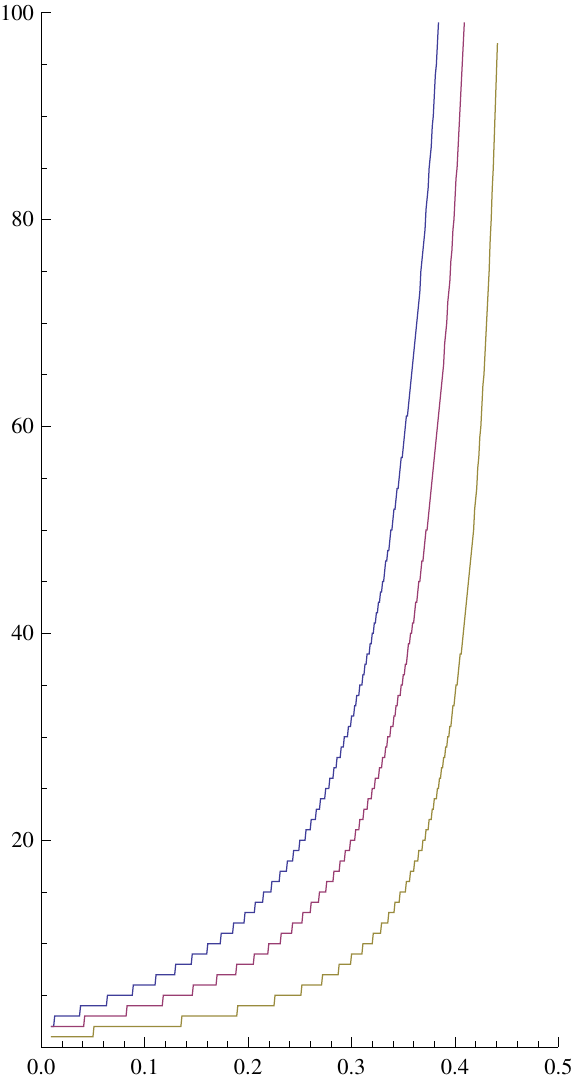}\\
(a)&(b)
\end{array}
\]
\caption{The number of confirmations required to keep the probability of success low, as a function of the attacker's hashrate, for various values of the target probability: 10\% (yellow), 1\% (purple) and 0.1\% (blue). The graph is shown in two different vertical scales.}
\label{fig:confirms}
\end{figure}

In \autoref{fig:confirms}, we show for three different probability targets (10\%, 1\% and 0.1\%), what is the number of confirmations required to keep the success rate below this target, as a function of the attacker's hashrate. This is shown in two different scales for the number of confirmations.

For example, if the attacker's hashrate is 10\% of the total network hashrate (0.1 on the horizontal axis), 2 confirmations are required to keep the success rate below 10\%, 4 confirmations are needed to have it less than 1\%, and 6 confirmations are necessary to decrease the probability of success below 0.1\%.

We can see that, for any given probability, as the attacker's hashrate approaches half of the total network hashrate, the number of confirmations required approaches infinity; for $q\ge0.5$, no amount of confirmations will suffice to decrease the probability, as it is always 100\%.

In \autoref{tbl:prob}, we see a detailed tabular reference for the probability of success, for different values of $q$ and $n$.

\begin{table}[h]
{\scriptsize
\begin{tabular}{r|rrrrrrrrrr}
q&1&2&3&4&5&6&7&8&9&10\\
\hline
2\%&4\%& 0.237\%& 0.016\%& 0.001\%& $\approx0$& $\approx0$& $\approx0$& $\approx0$& $\approx0$& $\approx0$\\
4\%&8\%& 0.934\%& 0.120\%& 0.016\%& 0.002\%& $\approx0$& $\approx0$& $\approx0$& $\approx0$& $\approx0$\\
6\%&12\%& 2.074\%& 0.394\%& 0.078\%& 0.016\%& 0.003\%& 0.001\%& $\approx0$& $\approx0$& $\approx0$\\
8\%&16\%& 3.635\%& 0.905\%& 0.235\%& 0.063\%& 0.017\%& 0.005\%& 0.001\%& $\approx0$& $\approx0$\\
10\%&20\%& 5.600\%& 1.712\%& 0.546\%& 0.178\%& 0.059\%& 0.020\%& 0.007\%& 0.002\%& 0.001\%\\
12\%&24\%& 7.949\%& 2.864\%& 1.074\%& 0.412\%& 0.161\%& 0.063\%& 0.025\%& 0.010\%& 0.004\%\\
14\%&28\%& 10.662\%& 4.400\%& 1.887\%& 0.828\%& 0.369\%& 0.166\%& 0.075\%& 0.034\%& 0.016\%\\
16\%&32\%& 13.722\%& 6.352\%& 3.050\%& 1.497\%& 0.745\%& 0.375\%& 0.190\%& 0.097\%& 0.050\%\\
18\%&36\%& 17.107\%& 8.741\%& 4.626\%& 2.499\%& 1.369\%& 0.758\%& 0.423\%& 0.237\%& 0.134\%\\
20\%&40\%& 20.800\%& 11.584\%& 6.669\%& 3.916\%& 2.331\%& 1.401\%& 0.848\%& 0.516\%& 0.316\%\\
22\%&44\%& 24.781\%& 14.887\%& 9.227\%& 5.828\%& 3.729\%& 2.407\%& 1.565\%& 1.023\%& 0.672\%\\
24\%&48\%& 29.030\%& 18.650\%& 12.339\%& 8.310\%& 5.664\%& 3.895\%& 2.696\%& 1.876\%& 1.311\%\\
26\%&52\%& 33.530\%& 22.868\%& 16.031\%& 11.427\%& 8.238\%& 5.988\%& 4.380\%& 3.220\%& 2.377\%\\
28\%&56\%& 38.259\%& 27.530\%& 20.319\%& 15.232\%& 11.539\%& 8.810\%& 6.766\%& 5.221\%& 4.044\%\\
30\%&60\%& 43.200\%& 32.616\%& 25.207\%& 19.762\%& 15.645\%& 12.475\%& 10.003\%& 8.055\%& 6.511\%\\
32\%&64\%& 48.333\%& 38.105\%& 30.687\%& 25.037\%& 20.611\%& 17.080\%& 14.226\%& 11.897\%& 9.983\%\\
34\%&68\%& 53.638\%& 43.970\%& 36.738\%& 31.058\%& 26.470\%& 22.695\%& 19.548\%& 16.900\%& 14.655\%\\
36\%&72\%& 59.098\%& 50.179\%& 43.330\%& 37.807\%& 33.226\%& 29.356\%& 26.044\%& 23.182\%& 20.692\%\\
38\%&76\%& 64.691\%& 56.698\%& 50.421\%& 45.245\%& 40.854\%& 37.062\%& 33.743\%& 30.811\%& 28.201\%\\
40\%&80\%& 70.400\%& 63.488\%& 57.958\%& 53.314\%& 49.300\%& 45.769\%& 42.621\%& 39.787\%& 37.218\%\\
42\%&84\%& 76.205\%& 70.508\%& 65.882\%& 61.938\%& 58.480\%& 55.390\%& 52.595\%& 50.042\%& 47.692\%\\
44\%&88\%& 82.086\%& 77.715\%& 74.125\%& 71.028\%& 68.282\%& 65.801\%& 63.530\%& 61.431\%& 59.478\%\\
46\%&92\%& 88.026\%& 85.064\%& 82.612\%& 80.480\%& 78.573\%& 76.836\%& 75.234\%& 73.742\%& 72.342\%\\
48\%&96\%& 94.003\%& 92.508\%& 91.264\%& 90.177\%& 89.201\%& 88.307\%& 87.478\%& 86.703\%& 85.972\%\\
50\%&100\%& 100\%& 100\%& 100\%& 100\%& 100\%& 100\%& 100\%& 100\%& 100\%
\end{tabular}
}
\caption{The probability of a successful double spend, as a function of the attacker's hashrate $q$ and the number of confirmations $n$.}
\label{tbl:prob}
\end{table}

With this more detailed analysis, we can note the following observations:
\begin{itemize}
\item{Successful double-spending is possible with any attacker hashrate. There is no need for a majority to carry out this attack.}
\item{Waiting for more confirmations exponentially decreases the probability of double-spending success. The decay rate depends on the attacker's relative hashrate.}
\item{The probability of success depends on the number of confirmations and \emph{not} on the amount of time waited. An alternative network with a different time constant $T_0$ can thus obtain more security with a given amount of wait time.}
\item{The time constant might be relevant, if we assume that the attacker cannot sustain his hashrate for a prolonged period of time. In this case, even majority hashrate does not guarantee success, as the attack could take more time than available. However, it does not seem likely that an attacker would go through the trouble of obtaining enormous computational resources and only maintain them for a time short enough to make this consideration relevant.} 
\item{No amount of confirmations will reduce the success rate to $0$.}
\item{If the attacker controls more hashrate than the honest network, no amount of confirmations will reduce the success rate below $100\%$.}
\item{There is nothing special about the default, often-cited figure of 6 confirmations. It was chosen based on the assumption that an attacker is unlikely to amass more than 10\% of the hashrate, and that a negligible risk of less than 0.1\% is acceptable. Both these figures are arbitrary, however; 6 confirmations are overkill for casual attackers, and at the same time powerless against more dedicated attackers with much more than 10\% hashrate.}
\end{itemize}

\section{Economics of double-spending}
The statistics of the previous section are directly relevant to merchants in an ``adversarial'' setting, who assume their customer is an attacker and want to make sure he will fail in his attack; they are overly conservative for a more realistic setting, where most customers are not attackers, and attackers will only target the merchant if it is profitable for them.

In this section we will assume that $q<p$. Otherwise, all bets are off with the current Bitcoin protocol. An attacker with majority hashrate can not only double-spend any transaction at no cost (above the cost of running the mining hardware normally); he can reject all blocks which are not his own, earning the entire coin generation reward himself. He can also exclude all transactions, disrupting the operation of the payment network. The honest miners, who no longer receive any rewards, would quit due to lack of incentive; this will make it even easier for the attacker to maintain his dominance. This will cause either the collapse of Bitcoin or a move to a modified protocol. As such, this attack is best seen as an attempt to destroy Bitcoin, motivated not by the desire to obtain Bitcoin value, but rather wishing to maintain entrenched economical systems or obtain speculative profits from holding a short position.

The exact economics of double-spending is of course highly complex; we will analyze a simple model that aims to highlight the key moving parts.

\begin{enumerate}
\item{A double-spending attack can be carried against more than one merchant. Payments can be simultaneously sent to $k$ different merchants, with the same branch invalidating all of them.}
\item{From each merchant products will be purchased with a cost of $v$, which we assume is equal for all merchants. Even if this does not strictly hold, an ``effective'' value of $v$ can be used in the model.}
\item{The products bought may not be completely liquid, and their value to the attacker may be less than $v$. We will denote their value by $\alpha v$ for some $0<\alpha\le1$. Again, $\alpha$ is assumed to be fixed or at least typical.}
\item{There is a tradeoff between $k$ and $\alpha$. There are only so many opportunities to extract value with sufficient anonymity to avoid prosecution. As the attacker tries to target more merchants, he will have to settle for decreasingly liquid assets. In-person transactions put timing constraints which again make it more difficult to target other merchants simultaneously.}
\item{There is also a tradeoff between $k$ and $v$, as $kv$ must be less than the attacker's total available capital. However, this is less of a concern, as the primary costs of the attack are likely to be loss of value and the block rewards, not the time value of money. The main limitation on $v$ is that higher values will cause the merchant to protect himself by waiting for more confirmations.}
\item{If the attack succeeds, all blocks found during the process will be valid, and the attacker will get for them the same reward as if he had not committed an attack (or possibly more). If the attack fails, and the attacker has found $o$ blocks during it, each with a block reward of $B$, those blocks will be rejected and the attacker will lose a total value of $oB$.}
\item{The attacker will obtain $k\alpha v$ worth of goods whether the attack succeeds or fails. It it fails, however, he will have to pay out $kv$.}
\end{enumerate}
We can see that, by performing an attack, the attacker will always gain $k\alpha v$, but if he fails, which happens with probability $1-r$ (a function of $n$ and $q$), he loses $kv+oB$. This means that his average profit is:
\[k\alpha v - (1-r)(kv+oB)=kv(\alpha+r-1)-(1-r)oB.\]
For the attack to be profitable, this has to be positive, meaning that
\[v>\frac{(1-r)oB}{k(\alpha+r-1)}\]
As long as the merchants make sure that this is not the case, they are safe because it would not be economical to carry out an attack.

We will assume for simplicity that $o=20$, meaning the attacker gives up after finding 20 blocks without successfully catching up, and that this does not impact his probability of success. This is a significant simplification; a more accurate model would determine an optimal stopping point by considering the reward of possibly completing the attack versus the risk of compounding the losses, and use the average blocks found until reaching that point. However, this assumption still preserves the spirit of the results.

This means that discouraging an attack requires that
\[v\le\frac{20(1-r)B}{k(\alpha+r-1)}\]

Unlike the adversarial setting, here increasing the time constant $T_0$ slightly improves security with a given number of confirmations, as it increases the effective value of $B$, and hence the maximal safe value. However, $r$ is the more dominant factor (because of its exponential dependence on the number of confirmations), and hence increasing $T_0$ worsens the security for a given wait time.

Modeling the practical values of $k$ and $\alpha$ is difficult, but we will assume that the best configuration the attacker manages to achieve is $\alpha=1,\ k=5$. Also assuming that $B = 25 \text{BTC}$, we find that if a merchants wait for $n$ confirmations, they are safe from double-spending as long as the value $v$ of each transaction satisfies
\begin{equation}
v\le\frac{100(1-r)}{r}\ \text{BTC}=100\left(\frac1r-1\right)\ \text{BTC}.
\label{eqn:value}
\end{equation}
The merchant would do well, however, to use an even lower value than this, to correct for the behavior of other merchants and any inaccuracies in the model. $r$ itself is a function of $n$ and $q$, given in \autoref{eqn:prob}.

A more accurate model might take into account that generating blocks costs more to the attacker than their reward, and that he would not have mined them at all (or procured the necessary computing resources) if he did not want to double-spend. Such a model could obviate the need to choose a value for $q$, by posing limits on the hashrate an attacker would obtain to perform attacks of a given value. However, once the focus of the security is the cost of neutrally mining, the number of confirmations required becomes linear, not logarithmic, in the transaction value; this is very poor security, hence in a situation where this is relevant, we have already lost anyway.

The maximal safe transaction values, for different values of the number of confirmations $n$ and attacker hashrate $q$, are given in \autoref{tbl:value}, based on \autoref{eqn:value}. These values should be taken with a grain of salt, because of the many modeling assumptions made.

\begin{table}[h]
{\footnotesize
\begin{tabular}{r|rrrrrrrrrr}
q&1&2&3&4&5&6&7&8&9&10\\
\hline
2\%& 2400& 42K& 644K& 9370K& $\approx\infty$& $\approx\infty$& $\approx\infty$& $\approx\infty$& $\approx\infty$& $\approx\infty$\\
4\%& 1150& 10K& 82K& 615K& 4437K& $\approx\infty$& $\approx\infty$& $\approx\infty$& $\approx\infty$& $\approx\infty$\\
6\%& 733& 4722& 25K& 127K& 626K& 3018K& 14M& $\approx\infty$& $\approx\infty$& $\approx\infty$\\
8\%& 525& 2650& 10K& 42K& 159K& 588K& 2144K& 7749K& $\approx\infty$& $\approx\infty$\\
10\%& 400& 1685& 5741& 18K& 56K& 168K& 503K& 1486K& 4361K& 12M\\
12\%& 316& 1158& 3391& 9212& 24K& 62K& 157K& 396K& 990K& 2460K\\
14\%& 257& 837& 2172& 5200& 11K& 27K& 60K& 132K& 290K& 632K\\
16\%& 212& 628& 1474& 3178& 6580& 13K& 26K& 52K& 102K& 200K\\
18\%& 177& 484& 1043& 2061& 3901& 7202& 13K& 23K& 42K& 74K\\
20\%& 150& 380& 763& 1399& 2453& 4190& 7039& 11K& 19K& 31K\\
22\%& 127& 303& 571& 983& 1615& 2582& 4053& 6288& 9671& 14K\\
24\%& 108& 244& 436& 710& 1103& 1665& 2467& 3608& 5229& 7525\\
26\%& 92& 198& 337& 523& 775& 1113& 1570& 2182& 3005& 4106\\
28\%& 78& 161& 263& 392& 556& 766& 1035& 1377& 1815& 2372\\
30\%& 66& 131& 206& 296& 406& 539& 701& 899& 1141& 1435\\
32\%& 56& 106& 162& 225& 299& 385& 485& 602& 740& 901\\
34\%& 47& 86& 127& 172& 221& 277& 340& 411& 491& 582\\
36\%& 38& 69& 99& 130& 164& 200& 240& 283& 331& 383\\
38\%& 31& 54& 76& 98& 121& 144& 169& 196& 224& 254\\
40\%& 25& 42& 57& 72& 87& 102& 118& 134& 151& 168\\
42\%& 19& 31& 41& 51& 61& 70& 80& 90& 99& 109\\
44\%& 13& 21& 28& 34& 40& 46& 51& 57& 62& 68\\
46\%& 8& 13& 17& 21& 24& 27& 30& 32& 35& 38\\
48\%& 4& 6& 8& 9& 10& 12& 13& 14& 15& 16\\
50\%& 0& 0& 0& 0& 0& 0& 0& 0& 0& 0
\end{tabular}
}
\caption{The maximal safe transaction value, in BTC, as a function of the attacker's hashrate $q$ and the number of confirmations $n$.}
\label{tbl:value}
\end{table}

\section{Conclusions}
In this paper we have explained the basic structure of the Bitcoin blockchain, the protection it gives against double-spending, and the ways in which this protection can be undermined. We have derived the probability for a successful double-spend, and tabulated it in various ways. We have also briefly discussed the conditions in which a double-spending attack can be economical, and hence likely. In so doing we have dispelled some popular myths, such as the absolute security believed to be granted by waiting for 6 confirmations, or the length of time waited (as opposed to the number of confirmations in terms of discrete blocks) as an allegedly major factor in determining security.

\newpage
\bibliographystyle{plain} % IEEEtr, plain, abbrv, alpha
\bibliography{Doublespend}
\end{document}